\documentclass[twocolumn,prb,10pt,showpacs,aps,footinbib]{revtex4}
\usepackage{graphicx}
\usepackage{color}
\usepackage{multirow}
\usepackage{parskip}
\usepackage{bm}

\begin{document}

\title{{$\bm{GW}$} quasiparticle band gap of the hybrid organic-inorganic perovskite CH$_3$NH$_3$PbI$_3$:
\vspace{0.1cm}\\
Effect of spin-orbit interaction, semicore electrons, and self-consistency} 

\author{Marina R. Filip}
\author{Feliciano Giustino}
\affiliation{Department of Materials, University of Oxford, Parks Road,
Oxford OX1 3PH, United Kingdom}

\pacs{
71.15.Mb, 
71.20.Nr, 
71.70.Ej, 
71.15.Qe  
}

\begin {abstract}
We study the quasiparticle band gap of the hybrid organic-inorganic lead halide 
perovskite CH$_3$NH$_3$PbI$_3$, using many-body perturbation theory based on the $GW$
approximation. We perform a systematic analysis of the band gap sensitivity to relativistic
spin-orbit effects, to the description of semicore Pb-5$d$ and I-4$d$ electrons,
and to the starting Kohn-Sham eigenvalues. We find that the inclusion of semicore
states increases the calculated band gap by 0.2~eV, and self-consistency on 
the quasiparticle eigenvalues using a scissor correction increases the band gap
by 0.4~eV with respect to the $G_0W_0$ result. These findings allow us to resolve
an inconsistency between previously reported $GW$ calculations for CH$_3$NH$_3$PbI$_3$. 
Our most accurate band gap is 1.65~eV, and is in good agreement with the measured optical 
gap after considering a small excitonic shift as determined in experiments.
\end {abstract}

\maketitle

\section{Introduction}
The development of solar cells based on metal-halide perovskites has shown unprecedented 
progress since perovskite absorbers were first reported in 2009.\cite{Kojima2009} 
In less than five years the power conversion efficiency of perovskite cells 
has increased from 3.8\%\cite{Kojima2009} to the current record of 19.3\%.\cite{Zhou2014}
Photovoltaic devices based on methylammonium (CH$_3$NH$_3^+$, MA) lead halide, MAPbI$_3$
and MAPbBr$_3$, were originally fabricated using
a dye-sensitized solar cell architecture.\cite{ORegan1991} In this device setup the
electrons generated upon absorption of light in the perovskite are collected in a
TiO$_2$ electron acceptor, while the hole is regenerated by a redox couple in an 
electrolytic solution.\cite{Kojima2009} After replacing the liquid electrolyte
by the solid-state hole-transporter spiro-MeOTAD, the efficiency of these solar cells 
increased above 10\%.\cite{Kim2012, Lee2012} This breakthrough generated a surge of interest 
in perovskites within the solar cell community.\cite{Green2014,Park2014} Further increase 
in the efficiency was achieved by improving fabrication techniques,\cite{Heo2013,Burschka2013,
Zhou2014} and by exploring mixed metal-halide perovskites.\cite{Noh2013}  

In view of large-scale industrial manufacturing of perovskite-based solar cells, 
several avenues of development are actively being explored. Ambipolar charge transport 
in the perovskite layer was demonstrated by the successful developement 
of simple planar heterojunction solar cells.\cite{Ball2013, Docampo2013, Liu2013, Liu2014, 
Eperon2014-3, Chen2014, He2014, Bai2014, Gamliel2014} Control of the color of the 
devices\cite{Noh2013,Eperon2014-2, Kulkarni2014} as well as partial 
transparency\cite{Eperon2014-2, Roldan-Carmona2014-2} were also achieved. Flexible thin 
film perovskite solar cells have been fabricated via low-temperature 
processing.\cite{Docampo2013, Kumar2013, You2014, Roldan-Carmona2014-1, Liu2014} Toxicity concerns
have also been addressed by studying alternatives to Pb such as Sn-based mixed-halide 
perovskite solar cells.\cite{Noel2014,Hao2014}

\begin{figure}[b]
\begin{center}
\includegraphics[width=0.4\textwidth]{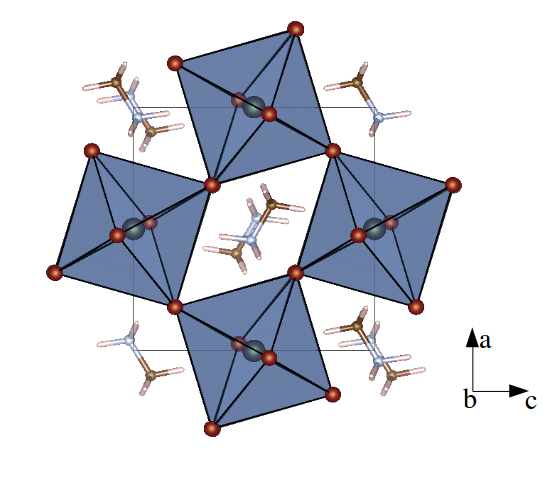}
\caption{
Polyhedral representation of the crystal structure of MAPbI$_3$. The Pb atoms are 
represented by the large blue spheres at the center of each octahedron. The small red 
spheres located at the shared corners of the octahedra are the I atoms. The MA cations are
shown using ball-and-stick models at the center of the cuboctahedral cavities (C is light blue,
N is brown, H is pink). The structure is viewed along the $b$ lattice vector.
\label{fig:fig1}}
\end{center}
\end{figure}

In parallel with these rapid technological advances, significant efforts are currently
being devoted to understanding the microscopic mechanisms which are responsible for 
the exceptional performance of these hybrid perovskites. MAPbI$_3$ belongs to the family 
of ABX$_3$ perovskites. The Pb and I atoms form a three-dimensional network of corner-sharing 
octahedra, with the Pb atoms occupying the center of each octahedron and the I atoms 
located at its corners. In this configuration the Pb-I network encloses a cuboctahedral cavity, 
with the MA cation located at its center (Fig.~\ref{fig:fig1}). The crystal structure of 
metal-halide perovskites is strongly dependent on temperature~\cite{Poglitsch1987, Baikie2013, 
Stoumpos2013} and undergoes two phase transitions. At low temperaure, the crystal structure 
of MAPbI$_3$ is orthorhombic and the MA cations have a well-defined orientation within 
the cuboctahedral cavity.\cite{Baikie2013,Stoumpos2013, Poglitsch1987}
Above 162~K the Pb-I network undergoes a phase transition to a tetragonal 
structure\cite{Baikie2013}, and above 327~K the system stabilizes in its most symmetric 
cubic perovskite structure.\cite{Baikie2013} As the temperature increases 
the MA cation becomes orientationally disordered. The structure of the Pb-I network 
in MAPbI$_3$ has been studied in detail for each phase using temperature dependent 
X-ray diffraction (XRD).\cite{Baikie2013, Stoumpos2013} These studies also noted structural 
anomalies relating to the position of the apical I atoms, as well as possible intermediate 
phases.\cite{Baikie2013,Stoumpos2013}

Given the uncertainties in the structure of the room-temperature tetragonal phase,
in this work we focus on the low-temperature orthorhombic structure of MAPbI$_3$.
The space group of this system is \emph{Pnma} and the orientation of the MA cations is well
understood.\cite{Poglitsch1987, Baikie2013, Stoumpos2013} Accurate experimental measurements 
of the optical band gap of the low-temperature phase of MAPbI$_3$ have been reported
for temperatures down to 4~K.\cite{Dinnocenzo2014}

The optical band gap of metal-halide perovskites can be broadly tuned by manipulating their
chemical composition and crystal structure.\cite{Knutson2005,Baikie2013,Stoumpos2013,
Eperon2014-1, Amat2014, Choi2014, Pellet2014, Filip2014, Kulkarni2014} 
The gap can be tuned from 1.17~eV to 1.55~eV by controlling the mixing fraction
of Pb and Sn in MAPb$_x$Sn$_{1-x}$I$_3$.\cite{Stoumpos2013, Hao2014-2} Perovskites
based on alternative cations such as Cs and formamidinium, HC(NH$_2$)$_2^+$, or on mixes 
of different cations showed a tunable optical onset, spanning a range of 
0.25~eV.\cite{Eperon2014-1,Choi2014,Pellet2014} Recently our computational study 
of the steric effect of the cation highlighted that the band gap could be finely tuned 
over a wide range by considering new cations not studied hitherto.\cite{Filip2014}
The band gap can also be tuned by exchanging the halide for Cl or Br.\cite{Eperon2014-2, 
Kulkarni2014, Noh2013} This possibility has led to band gap tunability in the range
1.5~eV to 2.3~eV.

MAPbI$_3$ is a semiconducting material, with an optical band gap of $\sim$1.6~eV at room 
temperature.\cite{Baikie2013} Optical absorption spectra measured as a function of
temperature identified a weakly bound exciton, with a binding energy estimated in 
the range $55\pm 20$~meV.\cite{Dinnocenzo2014} By applying the law of mass action 
for Wannier excitons the authors of Ref.~\onlinecite{Dinnocenzo2014} concluded that
at room temperature free carriers dominate over excitons in MAPbI$_3$. In addition
it is now understood that both electrons and holes have very long diffusion lenghts,
reaching up to 1~$\mu$m.\cite{Stranks2013, Wehrenfenig2013, Xing2013, Marchioro2014, 
Oga2014} Therefore it appears that MAPbI$_3$ perovskite share important similarities 
with standard tetrahedral semiconductors such as GaAs, and this may partly account 
for their extraordinary performance.\cite{Dinnocenzo2014, Stranks2014, Bergmann2014, 
Filippetti2014}

Computational studies of metal-halide perovskites within density functional theory 
(DFT)\cite{Hohenberg1964} have explored a variety of electronic and structural 
properties. In particular, perovskites with mixed I and Cl halides have been investigated,
highlighting the role of Cl incorporation in the structural and electronic properties of bulk 
MAPbI$_{3-x}$Cl$_x$,\cite{Mosconi2013} and the interface between TiO$_2$ and MAPbI$_{3-x}$Cl$_x$ 
or MAPbI$_3$.\cite{Lindblad2014,Mosconi2014,Roiati2014} The role of the cation as a spacer 
in the Pb-I network was established in computational studies of metal-halide 
perovskites.\cite{Amat2014, Filip2014} In particular, the size of the cation is understood
to determine the volume and the structure of the cuboctahedral cavity.\cite{Boriello2008,
Brivio2013,Amat2014,Filip2014} The effect of the orientation of MA has been 
studied in relation to UV/Vis absorption and Raman spectra,\cite{Grancini2014, Quarti2014} 
and a link with the hysteresis effect observed in the electrical measurements of 
perovskite-based solar cells has been suggested.\cite{Frost2014-1}  

DFT calculations of the band structure of MAPbI$_3$ revealed that the fundamental band gap 
is direct and located at the zone center. The valence band top and the conduction band
bottom are predominantly of I-$5p$ and Pb-$6p$ character, respectively.\cite{Chang2004,Baikie2013,
Filip2014,Lindblad2014,Filippetti2014} The electronic states associated with the MA cation are located more than
5~eV away from the band edges, therefore they are not directly involved in the
optical absorption onset.\cite{Filip2014, Lindblad2014} The band gap calculated using scalar relativistic DFT
is surprisingly close to the measured optical gap.\cite{Umari2014, Brivio2013, Chang2004, 
Even2013-1, Even2013-2} However, after incorporarting spin-orbit interactions using 
fully relativistic DFT calculations, the gap becomes about 1~eV smaller than 
in experiments.\cite{Even2013-1, Even2013-2}

The significant discrepancy between fully-relativistic DFT band gaps and experimental
optical gaps provided motivation for carrying out more sophisticated $GW$ quasiparticle 
calculations.\cite{Hedin} So far two studies addressed the quasiparticle band gap 
of MAPbI$_3$ within the $GW$ approximation, using different approaches.\cite{Umari2014, Brivio2014} 
Ref.~\citenum{Umari2014} reported a quasiparticle band gap of 1.67~eV obtained within
the $G_0W_0$ approximation. In this work the authors used a scalar relativistic
approximation for the screened Coulomb interaction $W_0$, and spin-orbit effects
were included only in the Green's function $G_0$ as a correction term.
Ref.~\citenum{Brivio2014} reported a $G_0W_0$ quasiparticle gap of 1.27~eV
when spin-orbit coupling was fully included in the calculation of both $G_0$ and $W_0$.
Ref.~\citenum{Brivio2014} also calculated the quasiparticle band gap using the
'quasiparticle self-consistent $GW$' (QS$GW$) approach.\cite{Schilfe2006} Surprisingly they
report a QS$GW$ band gap of 1.67~eV which matches the $G_0W_0$ results obtained in Ref.~\citenum{Umari2014}.

The present work was motivated by the difficulty in reconciling the findings 
of Refs.~\citenum{Umari2014} and \citenum{Brivio2014}. Our aim was to understand the 
discrepancy between the $G_0W_0$ band gaps calculated in Refs.~\citenum{Umari2014, Brivio2014}, 
and also to perform a sensitivity analysis of the band gap on the various approximations involved
in $GW$ calculations. In particular we investigated the effect of spin-orbit coupling
both in the DFT and in the $GW$ band gaps, by comparing scalar relativistic and 
fully relativistic calculations. We studied the effect of the semicore $d$ electrons 
of Pb and I on the calculated $GW$ band gaps. And we checked the numerical convergence 
of the band gap with respect to the cutoff on the number of empty states used in the 
Green's function and in the polarizability, with respect to the planewaves cutoff 
of the polarizability, and the Brillouin zone sampling.

Importantly we found that our best-converged, fully-relativistic $G_0W_0$ calculations 
yield a band gap which is about 0.4~eV smaller than the measured optical gap.
In order to improve the agreement with experiment, we tested whether a simple self-consistent
scissor-correction approach would increase the band gap, in the same spirit as $GW$+U 
calculations.\cite{Kioupakis2008,Patrick2012} Using this procedure
we obtained a band gap of 1.65~eV, in very good agreement with experiment and with the
calculations of Refs.~\citenum{Umari2014, Brivio2014}. We rationalize this result
by arguing that, in the case of MAPbI$_3$, the self-consistent scissor correction 
is effectively equivalent to performing quasi-particle self-consistent $GW$ as in 
Refs.~\citenum{Brivio2014}, or to using the scalar relativistic $W$ as in 
Refs.~\citenum{Umari2014}.

The manuscript is organized as follows. In Sec.~\ref{sec.methodology} we briefly summarize
the $GW$ formalism, and indicate how we perform self-consistent scissor calculations.
In Sec.~\ref{sec.computational} we describe our computational setup and analyze 
the numerical convergence of the band gap with empty states, planewaves cutoff, and
Brillouin zone sampling. Our main findings are presented in Sec.~\ref{sec.results}.
In Sec.~\ref{sec.discussion} we discuss our results in relation to experiments
and to the previous $GW$ calculations of Refs.~\citenum{Umari2014} and \citenum{Brivio2014}.
In Sec.~\ref{sec.conclusion} we offer our conclusions and identify important
avenues for future research.

\section{Methodology}\label{sec.methodology}

\subsection{Standard $\bm{G_0W_0}$ method}

Within $GW$ many-body perturbation theory the quasiparticle energies 
are written as:\cite{Hedin,Hybertsen,Arya1998} 
  \begin{equation}\label{eq:eq1}
  E_{n\mathbf{k}} = \varepsilon_{n\mathbf{k}}
                    + Z\left(\varepsilon_{n\mathbf{k}}\right)
                    \langle n\mathbf{k}|\hat{\Sigma}\left(\varepsilon_{n\mathbf{k}}\right)
                    -V_{xc}|n\mathbf{k}\rangle,
  \end{equation}
where $|n\mathbf{k}\rangle$ denotes a Kohn-Sham state with band index $n$ and crystal
momentum $\mathbf{k}$, and $\varepsilon_{n\mathbf{k}}$ its corresponding eigenvalue.
In Eq.~(\ref{eq:eq1}) $\hat{\Sigma}(\omega)$ is the self-energy at the energy $\omega$, 
$Z(\omega) = \big(1-\partial {\rm Re}\Sigma/\partial \omega)^{-1}$ is the quasiparticle 
renormalization, and $V_{xc}$ is the DFT exchange and correlation potential. 
The self-energy is calculated in the $G_0W_0$ approximation as $\hat{\Sigma} = 
i G_0W_0$.\cite{Hybertsen, Arya1998} 
Here $G_0$ is the non-interacting single particle Green's function, as obtained 
from the Kohn-Sham eigenstates.  
The screened Coulomb interaction is given by $W_0=\epsilon^{-1} v$, 
with $v$ the bare Coulomb interaction and $\epsilon^{-1}$ the inverse dielectric matrix.
The dielectric matrix is calculated in reciprocal space as:
  \begin{equation}\label{eq:eq2}
  \epsilon_{\mathbf{G}\mathbf{G'}}(\mathbf{q}, \omega) = 
      \delta_{\mathbf{G}\mathbf{G'}} - v(\mathbf{\mathbf{q}+\mathbf{G}})
     P_{\mathbf{G}\mathbf{G'}}(\mathbf{q}, \omega),
  \label{eq:2}
  \end{equation}
where $\mathbf{G}$ and $\mathbf{G'}$ are reciprocal lattice vectors, $\mathbf{q}$ is 
a wave vector in the first Brillouine zone, and $P_\mathbf{GG'}(\mathbf{q},\omega)$ is the 
polarizability in the random phase approximation (RPA).\cite{Adler1962, Wiser1963, Hybertsen}
In order to describe the frequency-dependence of the dielectric matrix we use the
Godby-Needs plasmon pole model.\cite{Godby, Onida, Arya1998} In this approximation 
the dielectric matrix is written as:
  \begin{equation}\label{eq:eq3}
  \epsilon_{\mathbf{GG'}}(\mathbf{q},\omega) = \delta_{\mathbf{GG'}}+\frac{\Omega_{\mathbf{GG'}}^
  2(\mathbf{q})}{\omega^2-[\omega_{\mathbf{GG'}}(\mathbf{q})-i\eta]^2},
  \end{equation}
where $\eta$ is a small constant. The plasmon-pole parameters $\Omega_{\mathbf{GG'}}$ and 
$\omega_{\mathbf{GG'}}$ are obtained by evaluating the RPA dielectric matrix at 
$\omega = 0$ and at $\omega=i\omega_{\rm p}$, $\omega_{\rm p}$ being the plasma 
frequency.\cite{Godby, Hybertsen, SaX, Arya1998} 
When performing convergence tests it is useful to partition the self-energy into an 
energy-independent term, the exchange self energy $\hat{\Sigma}_x$, and an energy-dependent 
term, the correlation self energy $\hat{\Sigma}_c$.

\subsection{Self-consistent scissor correction}\label{sec.theory-ssgw}

The $G_0W_0$ quasiparticle energies can be sensitive to the Kohn-Sham eigenvalues and 
eigenstates used as the starting point to calculate the perturbative corrections via 
Eq.~(\ref{eq:eq1}).\cite{Fuchs, Bruneval, Kioupakis2008,Patrick2012,Caruso2012, Caruso2013,
Schilfe2006} 
It is now well established that this sensititivy can be mitigated by employing a self-consistent 
approach in the many body perturbation theory problem, either by iteratively improving
$G$ and/or $W$, or by modifying the DFT starting point. Several approaches with varying
levels of sophistication have been proposed to this effect.\cite{Caruso2012, Caruso2013, 
Kioupakis2008,Patrick2012,Schilfe2006}

In this study we test the use of a simple scissor correction approach in order to study 
the impact of self-consistency on the quasiparticle band gap of MAPbI$_3$.
The scissor correction is particularly appropriate for MAPbI$_3$ since the optical
absorption onset results from transitions between two parabolic bands which are well
separated from other bands, therefore the effects of band mixing via off-diagonal
matrix elements of the self-energy are expected to be very small.

In this approach we first determine the $G_0W_0$ band gap correction $\Delta$ using
Eq.~(\ref{eq:eq1}). Subsequently we repeat the $GW$ calculation, after having applied
a scissor correction of magnitude $\Delta$ to the conduction bands. This procedure
is repeated until the correction $\Delta^{(i)}$ and $\Delta^{(i+1)}$ at two subsequent
iterations $i$ and $i+1$ are the same within a set tolerance.
The scissor correction modifies both the Green's function $G_0$ and the screened
Coulomb interaction $W_0$ via the RPA polarizability. For definiteness in the
following we will refer to this approach as `SS-$GW$' (self-consistent scissor $GW$).

The effect of the scissor correction can be seen as if obtained by adding a non-local
exchange and correlation potential to the orginal Kohn-Sham Hamiltonian, 
$\hat{V}_{\rm nl} = \Delta \, \hat{P}_c$, with $\hat{P}_c$ being the projector on the manifold 
of unoccupied states. In order to avoid double-counting this extra potential must be
removed from Eq.~(\ref{eq:eq1}); we obtain:
   \begin{equation}\label{eq:eq4}
   E_{n\mathbf{k}} = \varepsilon_{n\mathbf{k}}
                    + Z\left(\varepsilon_{n\mathbf{k}}\right)
                    \langle n\mathbf{k}|\hat{\Sigma}\left(\varepsilon_{n\mathbf{k}}\right)
                    -\left(V_{xc}+\hat{V}_{\rm nl}\right)|n\mathbf{k}\rangle.
  \end{equation}
In this form the analogy with other techniques, such as $GW$+U\cite{Kioupakis2008,Patrick2012}
or quasi-particle self-consistent $GW$\cite{Schilfe2006} becomes evident.
The main difference with more sophisticated approaches is that Eq.~(\ref{eq:eq4}) is
extremely easy to implement as a post-processing operation, and its effect on the
quasiparticle corrections is transparent.

We note in passing that the scissor-corrected $GW$ self-energy has a well-defined upper bound.
In fact, if we consider the limit of large $\Delta$ we find that the RPA dielectric
matrix becomes the identity, therefore the self-energy reduces to $G_0v$. After separating
the Green's function into its analytic and non-analytic components and noting that 
the frequency convolution of the analytic part vanishes,\cite{Giustino2010} we obtain that
for $\Delta\rightarrow\infty$ the self-energy and the renormalization factor 
tend to $\hat{\Sigma} = \hat{\Sigma}_x$ and $Z=1$, respectively.

This indicates that the scissor correction only affects the correlation self-energy $\Sigma_c$,
and acts to improve the description of the screening via a modulation of the band gap. 

It goes without saying that one could replace $\hat{V}_{\rm nl}$ by more advanced
options, such as for instance hybrid functional with varying fractions of exchange.
\cite{Alkauskas2008,Noori2012} However here for the sake
of simplicity we do not explore these alternatives.

\section{Computational setup}\label{sec.computational}

\subsection{DFT calculations}\label{sec.computational.dft}

All DFT calculations are performed using the \texttt{Quantum ESPRESSO} software package\cite{QE} 
within the local density approximation (LDA).\cite{Perdew, Ceperley}
For C, N and H we use non-relativistic, norm-conserving von Barth-Car\cite{VBC} 
pseudopotentials from the \texttt{Quantum ESPRESSO} pseudopotential library. For Pb and I 
we generate two sets of Troullier-Martins\cite{TM} norm conserving, fully relativistic 
pseudopotentials using the \texttt{ld1.x} pseudopotential generation tool of the
\texttt{Quantum ESPRESSO} suite. We consider the following valence atomic configurations: 
Pb without (6$s^2$6$p^2$) or with semicore states (5$d^{10}$6$s^2$6$p^2$), 
I without (5$s^2$5$p^3$) or with semicore states (4$d^{10}$5$s^2$5$p^3$). 
The total energy is converged within 6~meV/atom using a planewaves kinetic energy cutoff 
of 100~Ry for the calculations without I-4$d$ semicore electrons. In order to achieve the
same level of convergence in the presence of I-4$d$ states we employ a cutoff of 150~Ry.
Self-consistent calculations are performed using a 6$\times$6$\times$6 $\Gamma$-centered 
Brillouine zone grid, comprising of 112 irreducible points.

In order to describe the low-temperature phase of MAPbI$_3$ we consider an orthorhombic
unit cell including 4 formula units (48 atoms). In all our calculations we use the 
atomic positions and lattice parameters obtained in our previous work\cite{Filip2014} 
by performing a complete structural optimization starting from the XRD data 
of Ref.~\citenum{Baikie2013}.

\subsection{$\bm{G_0W_0}$ calculations}

All the $GW$ calculations are performed using version 3.4.1 of the \texttt{Yambo} 
software package\cite{Yambo}, which includes relativistic spin-orbit coupling within
the $GW$ implementation.\cite{Sakuma2011, Arya2008, Arya2009}

{\it Plasmon pole approximation --}
In order to check whether the plasmon-pole approximation in Eq.~(\ref{eq:eq3}) is adequate 
for MAPbI$_3$ we check the sensitivity of the calculated band gap to the plasmon pole parameter 
$\omega_{\rm p}$. To this aim we perform $G_0W_0$ calculations using
$\omega_{\rm p}=$7, 14, 20, 27, and 34~eV. The corresponding variation of the gap
is found to be at most 10~meV, suggesting that the dynamical screening is
correctly described. These tests correspond to scalar-relativistc calculations
in the absence of semicore electrons, using 240~unuccupied states, kinetic energy cutoffs
of 136~eV and 54~eV for the exchange and correlation parts of the self-energy,
respectively, and an unshifted 2$\times$2$\times$2 Brillouin zone grid.

{\it Brillouin zone sampling --}
We test the convergence of the quasiparticle band gap with respect to the sampling
of the Brillouin zone used in the convolution of $G$ and $W$ by comparing the band
gaps obtained using unshifted 2$\times$2$\times$2, 3$\times$3$\times$3 and 4$\times$4$\times$4 
meshes (with 8, 14, and 36 irreducible ${\bf k}$-points, respectively). 
The band gap is found to change by less than 10~meV throughout. These tests were carried 
out using the same parameters reported in the previous paragraph.

\begin{figure*} 
\begin{center}
\includegraphics[width=\textwidth]{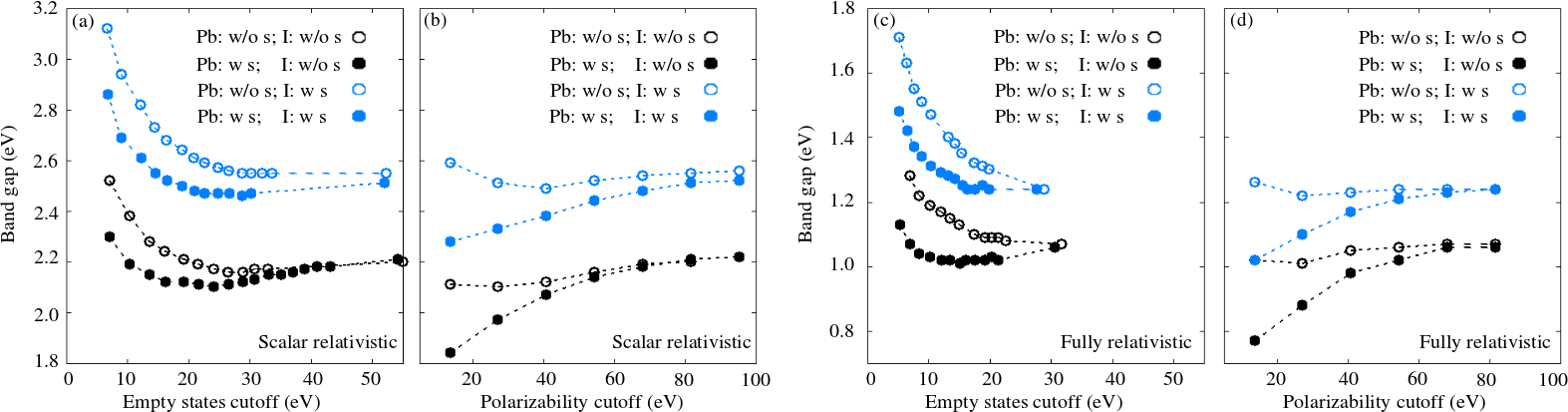}
\caption{
Convergence of the quasiparticle band gap of MAPbI$_3$ with respect to the energy
cutoff for unoccupied states [(a) and (c)], and with respect to the planewaves kinetic energy
cutoff for the polarizability [(b) and (d)]. Panels (a) and (b) refer to SR calculations,
while (c) and (d) are for FR band gaps. In each panel we show results for all four combinations 
of pseudopotentials considered here: Pb and I without semicore states (`w/o s', black empty 
circles), Pb with semicore (`w s') and I without semicore states (black filled circles), 
Pb without semicore and I with semicore states (blue empty circles), Pb and I with semicore 
states (blue filled circles). These data correspond to a 2$\times$2$\times$2 $\Gamma$-centered 
Brillouin zone mesh, and a 136~eV planewaves cutoff for the exchange self-energy. 
The planewaves cutoff for the polarizability in panels (a) and (c) is set to 82~eV, 
and the number of total bands in (b) and (d) is set to 1000.
\label{fig:fig2}}
\end{center}
\end{figure*}

{\it Unoccupied states --}
Since the planewaves kinetic energy cutoff for the polarizability and the number of
unoccupied states depend on whether we include semicore $d$ states in our calculations,
we study the convergence for all four combinations of pseudopotentials 
for Pb and I, as described in Sec.~\ref{sec.computational.dft}. For completeness we
carry out these tests both for scalar-relativistic (SR) and for fully relativistic (FR) 
calculations. In total we consider 8 scenarios, as shown in Fig.~\ref{fig:fig2}.

In order to determine optimal cutoff parameters we perform convergence tests for $\Sigma_x$ 
and $\Sigma_c$ separately. We found that a cutoff of 136~eV is sufficient to achieve 
convergence for $\Sigma_x$ in all cases. 

In the case of $\Sigma_c$ we test the convergence of the band gap with the number of empty 
states, for polarizability cutoffs ranging from 14 to 82~eV for all cases. We define
an energy cutoff for the unoccupied states using the energy of the highest state at $\Gamma$, 
relative to the valence band top. As a reference, the rightmost datapoints in 
Fig.~\ref{fig:fig2}(a) and (c) correspond to 1000~bands in total.

In the case of SR calculations the quasiparticle band gaps are converged with respect 
to the number of unoccupied states using a cutoff of 50~eV, corresponding to 1000~bands.
 
In the case of FR calculations the number of unoccupied states that need to be evaluated 
doubles with respect to SR calculations.
This can be seen in Fig.~\ref{fig:fig2}(a) and (c), where 1000 states correspond
to a cutoff of 30~eV and 55~eV in the FR and SR cases, respectively.
This poses in principle a limit on how far we can push the convergence of FR band gaps.
However from Fig.~\ref{fig:fig2} we can see that the SR and FR curves follow very similar
trends. This is expected given that the same pseudopotentials are used for both
sets of calculations. By comparing the convergence trends in the SR and FR cases 
we estimate that the fully-relativistic gaps calculated for the rightmost points 
in Fig.~\ref{fig:fig2}(c) are converged within less than 0.1~eV in all cases. 

{\it Polarizability cutoff --}
The convergence of the band gap with respect to the polarizability cutoff is shown in 
Fig.~\ref{fig:fig2}(b) and (d). The calculations are performed using 1000~bands in all cases. 
In order to contain the computational effort we tested cutoffs up to 95~eV in the SR case
and 82~eV in the FR case. In all cases cosidered the quasiparticle band gaps 
appear converged when using a polarizability cutoff of 82~eV. 

Having performed a detailed analysis of each convergence parameter, for the sake
of clarity and reproducibility in the remainder of the manuscript we report
quasiparticle band gaps calculated using the following setup:
planewaves kinetic energy cutoff for the exchange and the correlation self-energies:
136~eV and 82~eV, respectively; total number of bands: 1000; plasmon pole parameter:
$\omega_{\rm p}=27$~eV; Brillouin zone sampling: 2$\times$2$\times$2 unshifted mesh.

\section{Results}\label{sec.results}

\subsection{Effect of spin-orbit coupling}

Table~\ref{tb:tb1} shows a comparison between DFT/LDA and $G_0W_0$ calculations 
of the band gap of MAPbI$_3$ within both the scalar-relativistic and fully relativistic frameworks.
The spin-orbit coupling is seen to induce a 0.9~eV red-shift of the DFT band gap,
in agreement with previous reports.\cite{Even2013-1, Even2013-2} 

Moving on to the $GW$ quasiparticle corrections, we note that the red-shift induced by
the spin-orbit coupling increases sligthly and ranges between 1.1 and 1.3~eV.
Our results are in line with the $G_0W_0$ calculations of Ref.~\citenum{Brivio2014}, 
where a quasiparticle spin-orbit coupling correction of 1.5~eV was reported.

Table~\ref{tb:tb1} also shows that our calculated quasi-particle band gap including
spin-orbit corrections and semicore electrons is 1.24~eV. This value underestimates
the measured optical gap by approximately 0.4~eV,\cite{Baikie2013} suggesting that
a perturbative treatment of the quasiparticle corrections is insufficient in the
study of MAPbI$_3$. We will come back to this aspect in Sec.~\ref{sec.selfcon}.

Interestingly $G_0W_0$ calculations starting from the DFT/LDA states without including
spin-orbit effects yield a very large gap of 2.51~eV, therefore they 
overestimate the measured optical gap by as much as 0.9~eV (Table~\ref{tb:tb1}).
This observation reinforces the point on the importance of relativistic spin-orbit
corrections in this system.\cite{Even2013-1}

\begin{table*}
\begin{center}
\begin{tabular}{crrrrrrrrrrrrrrrrrrrrrrrr}
\hline
\hline
  \\[-8pt]
  && \multicolumn{11}{c}{DFT/LDA}& & \multicolumn{11}{c}{$G_0W_0$}\\
  \\[-8pt]
  \hline
  \\[-8pt]
  &&\hspace{0.2cm} Pb:& w/o s & & Pb:& w   s & & Pb:& w/o s & & Pb:& w s & \hspace{1cm} 
			& Pb:& w/o s & & Pb:& w s & & Pb:& w/o s & & Pb:& w s \\
  &&  I:& w/o s & &  I:& w/o s & &  I:& w s   & &  I:& w s &  
                        & I:& w/o s & &  I:& w/o s & &  I:& w s & &  I:& w s\\
  \\[-8pt]
  \hline
  \\[-8pt]
  \multirow{2}{*}{\vspace{-0.2cm}$\Gamma_v$}\hspace{0.5cm} & \multicolumn{1}{l}{SR} & \multicolumn{2}{r}{0.00} & & \multicolumn{2}{r}{0.00} & & \multicolumn{2}{r}{0.00} & & \multicolumn{2}{r}{0.00} &
                                    & \multicolumn{2}{r}{-0.30} & & \multicolumn{2}{r}{-0.31} & &\multicolumn{2}{r}{-0.45} & & \multicolumn{2}{r}{-0.53} \\
  \\[-8pt]
   & \multicolumn{1}{l}{FR} & \multicolumn{2}{r}{0.00} & & \multicolumn{2}{r}{0.00} & & \multicolumn{2}{r}{0.00} & & \multicolumn{2}{r}{0.00} &
                                    & \multicolumn{2}{r}{0.06} & & \multicolumn{2}{r}{0.09} & &\multicolumn{2}{r}{0.12} & & \multicolumn{2}{r}{0.14} \\
  \\[-8pt]
  \hline
  \\[-8pt]
  \multirow{2}{*}{\vspace{-0.2cm}$\Gamma_c$}\hspace{0.5cm} & \multicolumn{1}{l}{SR} & \multicolumn{2}{r}{1.43} & & \multicolumn{2}{r}{1.42} & & \multicolumn{2}{r}{1.51} & & \multicolumn{2}{r}{1.50} &
                                    & \multicolumn{2}{r}{1.90} & & \multicolumn{2}{r}{1.90} & &\multicolumn{2}{r}{2.10} & & \multicolumn{2}{r}{1.98} \\
  \\[-8pt]
  & \multicolumn{1}{l}{FR} & \multicolumn{2}{r}{0.54} & & \multicolumn{2}{r}{0.52} & & \multicolumn{2}{r}{0.60} & & \multicolumn{2}{r}{0.58} &
                                    & \multicolumn{2}{r}{1.13} & & \multicolumn{2}{r}{1.15} & &\multicolumn{2}{r}{1.36} & & \multicolumn{2}{r}{1.38} \\
  & \\[-8pt]
  \hline
  \\[-8pt]
  \multirow{3}{*}{\vspace{-0.5cm}Gap}\hspace{0.5cm} & \multicolumn{1}{l}{SR} & \multicolumn{2}{r}{1.43} & & \multicolumn{2}{r}{1.42} & & \multicolumn{2}{r}{1.51} & & \multicolumn{2}{r}{1.50} &
                                    & \multicolumn{2}{r}{2.20} & & \multicolumn{2}{r}{2.21} & &\multicolumn{2}{r}{2.55} & & \multicolumn{2}{r}{2.51} \\
  \\[-8pt]
  & \multicolumn{1}{l}{FR} & \multicolumn{2}{r}{0.54} & & \multicolumn{2}{r}{0.52} & & \multicolumn{2}{r}{0.60} & & \multicolumn{2}{r}{0.58} &
                                    & \multicolumn{2}{r}{1.07} & & \multicolumn{2}{r}{1.06} & &\multicolumn{2}{r}{1.24} & & \multicolumn{2}{r}{1.24} \\
  \\[-8pt]
  & \multicolumn{1}{l}{Difference} & \multicolumn{2}{r}{0.89} & & \multicolumn{2}{r}{0.90} & & \multicolumn{2}{r}{0.91} & & \multicolumn{2}{r}{0.92} &
                                    & \multicolumn{2}{r}{1.13} & & \multicolumn{2}{r}{1.15} & &\multicolumn{2}{r}{1.31} & & \multicolumn{2}{r}{1.27} \\
  \\[-8pt]
  \hline
  \hline
\end{tabular}
  \caption{\label{tb:tb1} 
  DFT/LDA Kohn-Sham energies and $G_0W_0$ quasiparticle energies of the valence band top 
  ($\Gamma_v$), conduction band bottom ($\Gamma_c$), and band gap of MAPbI$_3$.
  We report the results of scalar relativistic calculations (SR) and fully relativistic
  calculations (FR) for all combinations of Pb and I pseudopotentials considered in this
  work (`w s'/`w/o s' indicates that semicore electrons are included/not included).
  All values are in eV and are referred to the DFT valence band top.
  The last row reports the difference between the SR and the FR gap for each case considered.
  }
  \end {center}
  \end{table*}

\begin{figure*}[t!] 
\begin{center}
\includegraphics[width=\textwidth]{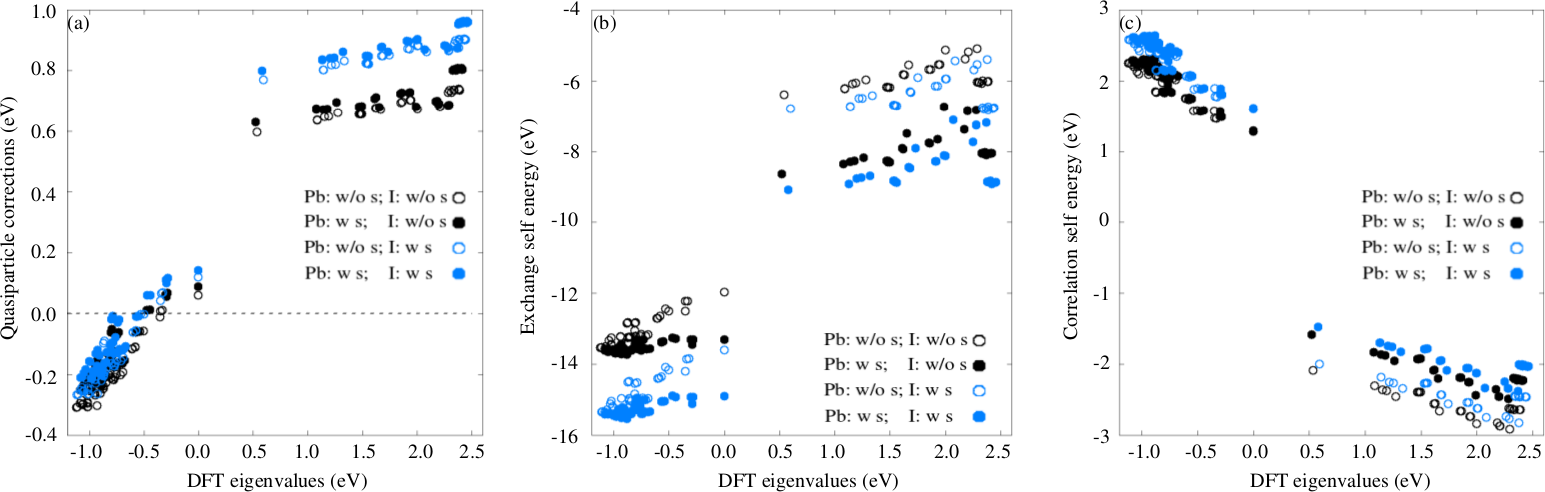}
\caption{
(a)Fully-relativistic quasiparticle corrections of the DFT/LDA band gap, 
(b) the exchange self-energy, and (c) the correlation self-energy of states
near the band edges and throughout the Brillouin zone, as a function of DFT/LDA eigenvalues.
The key to the symbols and the calculation parameters are the same as in 
Fig.~\protect\ref{fig:fig2}. 
\label{fig:fig3}}
\end{center}
\end{figure*}

\subsection{Effect of semicore electrons}\label{sec.semicore}

Table~\ref{tb:tb1} shows that the DFT/LDA band gap fo MAPbI$_3$ changes by less than 0.04~eV
upon the inclusion of the Pb-$5d$ and I-$4d$ semicore electrons in the valence. 
By contrast, the quasiparticle band gap appears sensitive to the Pb-$5d$ and I-$4d$ 
states: upon inclusion of these states the band gap increases by up to 0.17~eV in the
case of fully relativistic calculations. 
This trend can also be seen in the convergence study shown in Fig.~\ref{fig:fig2}.
A significant fraction of this shift arises from the inclusion of I-$4d$ states,
as can be seen in the fifth line of Table~\ref{tb:tb1}. The effect appears to be
mostly associated with a blue-shift of the conduction band bottom upon inclusion
of the I semicore electrons, as it is clearly seen in Fig.~\ref{fig:fig3}(a).

The effect of semicore electrons on the calculation of quasiparticle energies is 
well documented in literature.\cite{Rohlfing1995, Tiago2004, Filip2013} Usually
this effect results from an additional exchange contribution to the self-energy,
which results from the spatial overlap of diffuse semicore states with the valence states
at the band edges. At variance with the exchange self-energy, the correlation
part is less affected since the contribution of the semicore states is damped by
the energy separation between valence and semicore (Pb-$5d$ and I-$4d$ states lie
around 15~eV and 40~eV below the valence band top, respectively).

In Fig.~\ref{fig:fig3}(b) and (c) we can see that, in line with the above reasoning,
the effect of semicore states is most pronounced ($\sim$1~eV) in the exchange part 
of the self-energy, and somewhat smaller for the correlation part ($\sim$0.5~eV).
Focusing on the exchange part [Fig.~\ref{fig:fig3}(b)] we note that the bottom
of the conduction band is mostly affected by the Pb-$5d$ electrons. This is consistent
with the predominant Pb-$6p$ character of those states.\cite{Filip2014,Filippetti2014}
Similarly the corrections near the top of the valence band are affected 
both by Pb-$5d$ and I-$4d$ semicore states, in line with the observation that
those states carry both Pb-$6s$ and I-$5p$ character.\cite{Filip2014, Filippetti2014}

Taken together these results indicate that $GW$ quasiparticle corrections
are sensitive to the explicit inclusion of semicore electrons in the valence,
and this may result in band gap variations around 15\%.

As a side note we mention that in all the calculations discussed in this section
we obtained a quasiparticle renormalization factor of $0.80\pm 0.07$
(both for SR and FR calculations, and irrespective of the valence configuration).

\begin{figure}[t!] 
\begin{center}
\includegraphics[width=0.35\textwidth]{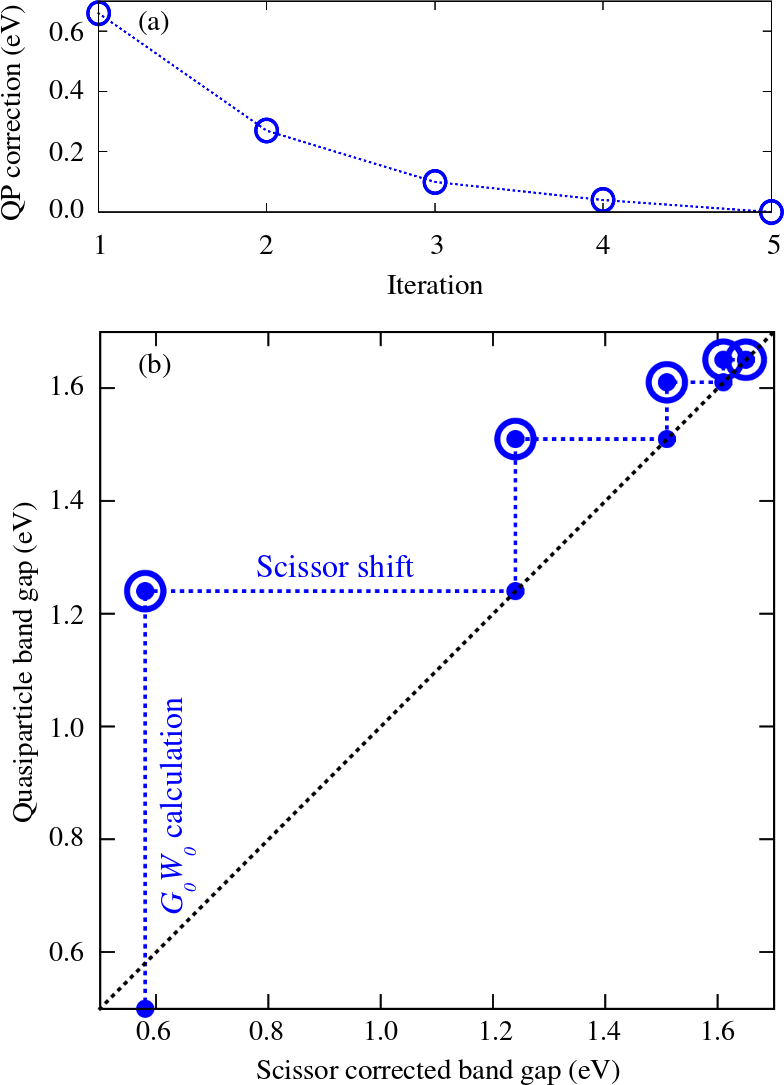}
\caption{
(a) Convergence of the self-consistent scissor correction to the band gap 
of MAPbI$_3$ with the number of iterations. The first iteration corresponds to
a $G_0W_0$ calculation using DFT/LDA eigenstates and eigenvalues (i.e.\ no scissor correction).
(b) Evolution of the quasiparticle band gap of MAPbI$_3$ throughout the SS-$GW$ 
iterative procedure. The blue dotted lines show the progression of the
calculation at each iteration: the vertical blue lines represent the $G_0W_0$ calculation 
performed at each iteration, while the horizontal blue lines are the scissor shifts
applied at each iteration. The iterative procedure stops when the $G_0W_0$ correction
vanishes, that is when the small and large circle do coincide on the black dotted line.
\label{fig:fig4}}
\end{center}
\end{figure}

\subsection{Effect of self-consistent scissor correction}\label{sec.selfcon}

For clarity in this section we only present results obtained from fully relativistic 
$GW$ calculations including both Pb-$5d$ and I-$4d$ semicore electrons in the valence.

Figure~\ref{fig:fig4}(a) shows the convergence of the quasiparticle correction to
the band gap at each iteration of the self-consistent scissor procedure described
in Sec.~\ref{sec.theory-ssgw}. The SS-$GW$ procedure converges after five iterations,
at which point the band gap correction becomes smaller than 10~meV.

A visual representation of the self-consistent procedure is provided in Fig.~\ref{fig:fig4}(b),
where we plot the quasiparticle band gap vs.\ the scissor-corrected gap of the previous
iteration. We note the close similarity of this plot with those obtained in previous 
$GW$+U calculations on different systems.\cite{Kioupakis2008,Patrick2012}

Our final self-consistent value of the quasiparticle band gap of MAPbI$_3$ is 1.65~eV. 

\section{Discussion}\label{sec.discussion}

Table~\ref{tb:tb3} reports our best results for the quasi-particle band gap of MAPbI$_3$.
For completeness we provide both scalar- and fully-relativistic $G_0W_0$ results
and the SS-$GW$ gap in the fully relativistic case.

In the same table we compare our results with the calculations of Refs.~\citenum{Umari2014} 
and \citenum{Brivio2014}, and with the measured optical gap reported in Ref.~\citenum{Dinnocenzo2014}.

\begin{table}[t!]
\begin{center} 
\begin{tabular}{ccccccccccc}
\hline 
\hline                                                                                        
  \\[-10pt] 
  & & \multicolumn{3}{c}{Present work} & & \multicolumn{3}{c}{Previous $GW$} & & \multicolumn{1}{c}{Experiment} \\
  \\[-10pt]
  & & \multicolumn{1}{c}{$G_0W_0$} & & \multicolumn{1}{c}{SS-$GW$} & & \multicolumn{1}{c}{$G_0W_0$} & & \multicolumn{1}{c}{QS$GW$} \\
  \\[-8pt]
\hline
  \\[-8pt]
SR& & 2.51     & &              & & 2.68$^\mathrm{a}$, 2.73$^\mathrm{b}$ & & & & \multirow{2}{*}{\vspace{-0.1cm}1.62-1.64$^\mathrm{c}$} \\
FR& & 1.24     & & 1.65         & & 1.67$^\mathrm{a}$, 1.27$^\mathrm{b}$ & & 1.67$^\mathrm{b}$    & &  \\
  \\[-8pt]
\hline
\hline
\multicolumn{9}{l}{$^\mathrm{a}$Ref.~\citenum{Umari2014}.}\\
\multicolumn{9}{l}{$^\mathrm{b}$Ref.~\citenum{Brivio2014}.}\\
\multicolumn{9}{l}{$^\mathrm{c}$Ref.~\citenum{Dinnocenzo2014}.}\\
\end{tabular}
\caption{
Comparison between our calculated quasiparticle band gaps of MAPbI$_3$,
obtained here within the $G_0W_0$ or the self-consistent scissor (SS-$GW$) approach, 
and previously reported values (in eV).
The measured optical gap is estimated by using the midpoint of the absorption onset
in the optical spectra of Ref.~\protect\citenum{Dinnocenzo2014}
[curves at 4.2~K in Fig.~1(a) and Fig.~1(b) of Ref.~\protect\citenum{Dinnocenzo2014}].
`$G_0W_0$' indicate standard perturbative $GW$ calculations, `QS$GW$' indicates 
quasiparticle self-consistent $GW$ calculations. `SR' and `FR' stand for scalar- and
fully-relativistic calculations, respectively.
}
\label{tb:tb3}
\end{center}
\end{table}

Our $G_0W_0$ SR band gaps are in good agreement with previous calculations,
and also overestimate the band gap with respect to experiment by up to 1~eV due to the
neglect of relativistic effects. FR calculations including spin-orbit coupling yield
much a improved $G_0W_0$ gap, in good agreement with the $G_0W_0$ value reported in
Ref.~\citenum{Brivio2014}. However, we should point out that this agreement is somewhat
fortuitous, since in Ref.~\citenum{Brivio2014} the quasi-particle renormalization
(which in our case is $Z=0.8$) was not taken into account. We speculate that the
agreement may be due to the fact that the authors of Ref.~\citenum{Brivio2014}
considered a cubic MAPbI$_3$ structure which is known to have a slightly
smaller gap than its orthorhombic counterpart. 

Our band gap obtained within the self consistent scissor approach (1.65~eV)
matches well the quasiparticle self-consistent $GW$ (QS$GW$) band gap reported 
in Ref.~\citenum{Brivio2014} (1.67~eV). This is not too surprising since
the energy-dependence of the quasiparticle corrections for MAPbI$_3$ is
very smooth, as in standard semiconductors like silicon [see Fig.~\ref{fig:fig3}(a)].
This should be an indication that our simple self-consistent scissor should
be able to mimic more elaborate self-consistent procedures.

The validity of the SS-$GW$ approach is especially interesting in the wider context of predicting
from first principles the band gap of a variety of hybrid perovskites. In fact this
approach is simple, transparent, and effective, and its computational cost
is essentially negligible once the initial $G_0W_0$ calculation has been
performed. This procedure could also be used as a way to gauge
the reliability of the perturbative treatment, by simply comparing
the initial $G_0W_0$ gap with the final SS-$GW$ gap: a significant discrepancy
between these values may signal the need to go beyond the standard $G_0W_0$ theory.

When we compare our results with the calculations of Ref.~\citenum{Umari2014} 
we find that our FR gap (1.24~eV) is about 0.4~eV smaller than their value (1.67~eV),
and that the latter result is very close to our SS-$GW$ gap.
This apparently puzzling situation can be rationalized by noting that in
Ref.~\citenum{Umari2014} the Green's function is fully relativistic, but the
screened Coulomb interaction is calculated at the scalar relativistic level.
This choice implies that the $W_0$ of Ref.~\citenum{Umari2014} was obtained
starting from a DFT/LDA gap which was artificially increased by the neglect
of spin-orbit effects. In doing so the $W_0$ of Ref.~\citenum{Umari2014}
was obtained using a DFT/LDA band gap which is close to our
scissor-corrected gap. In order words the choice of Ref.~\citenum{Umari2014}
of neglecting spin-orbit effects in $W_0$ acts as an `effective scissor
correction'. This observation nicely reconciles the apparent discrepancy
between the results of Ref.~\citenum{Umari2014} and Ref.~\citenum{Brivio2014}.

Moving on to a comparison with experiment, since we are studying the
low-temperature orthorhombic structure at $T=0$ we consider the gap measured 
in Ref.~\onlinecite{Dinnocenzo2014} at 4.2~K. 
The exciton binding energy estimated in Ref.~\onlinecite{Dinnocenzo2014}
is 55 ± 20 meV, and the measured optical gap is 1.62-1.64 eV.
These values are in very good agreement with our
calculated gap of 1.65 eV, especially if we account
for the width of the optical lineshape.

At this point we should mention that our calculations do not include the
effect of the zero-point renormalization induced by the electron-phonon
interaction.\cite{Allen-Heine1976} 
However, while these effects can be large in light elements
like carbon,\cite{Giustino2010prl, Marini2012,Gonze2014} in the case
of MAPbI$_3$ they are expected to be small due to the heavy masses of Pb and~I.

\section{Conclusions}\label{sec.conclusion}

In this manuscript we reported a systematic study of the quasi-particle band gap
of the hybrid organic-inorganic lead halide perovskite MAPbI$_3$ 
(MA = CH$_3$NH$_3^+$), using $GW$ many-body perturbation theory.

We placed an emphasis on the reliability and reproducibility of our calculations 
by providing comprehensive convergence tests, as well as a detailed assessment of
the accuracy of our calculations. 

We found that the explicit inclusion of semicore Pb-$5d$ and I-$4d$ electrons
in the valence manifold is important in order to obtain a quantitatively accurate 
band gap. In fact, the neglect of the semicore states can induce errors in the gap
of the order of 15\%.

We demonstrated that the converged $G_0W_0$ band gap including semicore electrons and spin-orbit
coupling underestimates significantly (by 0.4~eV) the experimental optical gap of MAPbI$_3$.
In order to address this deficiency we tested a simple self-consistent scissor correction
approach (SS-$GW$). We demonstrated that this simple method can achieve very good agreement
with experiment and with more elaborate self-consistent $GW$ schemes.

Our calculations allowed us to clarify the apparent discrepancy between two previous
$GW$ calculations of the band gap of MAPbI$_3$.\cite{Umari2014, Brivio2014} In particular 
we showed that the calculations of Ref.~\citenum{Umari2014}, by neglecting spin-orbit 
effects in the evaluation of the screened Coulomb interaction, effectively incorporate
an element of self-consistency. After taking this aspect into account our calculations
and those of Refs.~\citenum {Umari2014, Brivio2014} turn out to be in good agreement.

Using self-consistent scissor approach we obtained a best-estimate for the band gap 
(1.65~eV), which is very close to available experimental data (1.62-1.64~eV).

Our present findings suggest that the SS-$GW$ approach may find wider application
in the first-principles prediction of the band gap of related perovskites. This could be
important in view of developing automated high-throughput computational screening strategies
relying on $GW$ calculations.

Given the well known excitonic properties of MAPbI$_3$,\cite{Dinnocenzo2014} future work 
should focus on a systematic computational study of the optical absorption spectrum within 
the Bethe-Salpeter formalism. Furthermore it will be important to carefully assess the
impact of electron-phonon interactions on the electronic and optical properties of MAPbI$_3$.

It is hoped that the detailed analysis presented in this work will serve as a reliable 
starting point for future studies of the electronic and optical properties
of this exciting material and the wider family of metal-halide hybdrid perovskites.

\begin{acknowledgements}
This work was supported by the European Research Council (EU FP7 / ERC grant no. 239578),  
the UK Engineering and Physical Sciences Research Council (Grant No. EP/J009857/1) and  
the Leverhulme Trust (Grant RL-2012-001). Calculations were performed at the Oxford Supercomputing  
Centre and at the Oxford Materials Modelling Laboratory. All structural models were rendered using
VESTA\cite{VESTA}.
\end{acknowledgements}

\bibliography{bibliography}

\end{document}